# A decomposition of Fisher's information to inform sample size for developing fair and precise clinical prediction models – part 1: binary outcomes

Richard D Riley[1, 2*], Gary S Collins[3], Rebecca Whittle[1, 2], Lucinda Archer[1, 2], Kym IE Snell[1, 2],

Paula Dhiman[3], Laura Kirton,[4] Amardeep Legha,[1, 2] Xiaoxuan Liu,[2] Alastair Denniston,[2]

Frank E Harrell Jr,[5] Laure Wynants,[6,7] Glen P. Martin,[8] Joie Ensor[1, 2]

* Corresponding author:  r.d.riley@bham.ac.uk   @Richard_D_Riley

**Author details:**
[1] Institute of Applied Health Research, College of Medical and Dental Sciences, University of Birmingham, Birmingham, UK
[2] National Institute for Health and Care Research (NIHR) Birmingham Biomedical Research Centre, Birmingham, UK.
[3] Centre for Statistics in Medicine, Nuffield Department of Orthopaedics, Rheumatology and Musculoskeletal Sciences, University of Oxford, Oxford, OX3 7LD, UK
[4] Cancer Research UK Clinical Trials Unit, Institute of Cancer and Genomic Sciences, College of Medical and Dental Sciences, University of Birmingham, Birmingham, UK
[5] Department of Biostatistics, Vanderbilt University School of Medicine, Nashville TN, USA
[6] Department of Epidemiology, Care and Public Health Research Institute (CAPHRI), Maastricht University, Maastricht, The Netherlands;
[7] Department of Development and Regeneration, KU Leuven, Leuven, Belgium
[8] Division of Informatics, Imaging and Data Science, Faculty of Biology, Medicine and Health, University of Manchester, Manchester Academic Health Science Centre, Manchester, UK

**Funding**: This paper presents independent research supported by an EPSRC grant for 'Artificial intelligence innovation to accelerate health research' (number: EP/Y018516/1), and by the National Institute for Health and Care Research (NIHR) Birmingham Biomedical Research Centre at the University Hospitals Birmingham NHS Foundation Trust and the University of Birmingham. GSC is supported by Cancer Research UK (programme grant: C49297/A27294). PD is supported by Cancer Research UK (project grant: PRCPJT-Nov21\100021). LK is supported by an NIHR Doctoral Fellowship (NIHR303331) and by a core funding grant awarded to the Cancer Research UK Clinical Trials Unit by Cancer Research UK (CTUQQR-Dec22/100006). GPM and RDR are partially supported by funding from the MRC-NIHR Methodology Research Programme [grant number: MR/T025085/1]. Funding for FEH Jr. was provided by The Vanderbilt Institute for Clinical and Translational Research (VICTR) from a grant by the U.S. National Center for Advancing Translational Sciences (NCATS) Clinical Translational Science Award (CTSA) Program, Award Number 5UL1TR002243-03. RDR, GSC and AD are National Institute for Health and Care Research (NIHR) Senior Investigators.  The views expressed are those of the author(s) and not necessarily those of the NHS, the NIHR or the Department of Health and Social Care.

**Competing Interests**: RDR receives royalties for textbooks on Prognosis Research and IPD Meta-Analysis.

**Contributions:** RDR derived the mathematical solutions, initial software code, made the initial applied examples and wrote the paper. JE produced the pmstabilityss module to generalise the code and apply to the examples. All authors provided critical feedback at multiple stages that let to revision of the rationale, methods, applications, fairness checks, and text or figures in the paper. RDR revised the paper after feedback from authors.

**Data:** The kidney dataset was obtained from the Medical Information Mart for Intensive Care III (MIMIC-III) trial, which contains freely available and de-identified critical care data from the Beth Israel Deaconess Medical Center in Boston, Massachusetts, between 2001 and 2012. See: Johnson AE, Pollard TJ, Shen L, Lehman LW, Feng M, Ghassemi M, et al. MIMIC-III, a freely accessible critical care database. Sci Data. 2016;3:160035.

**Word count:** 8600

**Human Ethics and Consent to Participate declarations**: not applicable

**Abstract**

**Background**

When developing a clinical prediction model, the sample size of the development dataset is important. Smaller sample increase concerns of overfitting, instability, poor predictive performance and a lack of fairness. For models estimating the risk of a binary outcome, previous research has outlined minimum sample size calculations to minimise overfitting and precisely estimate the overall risk. However, even when meeting these criteria, the uncertainty (instability) in individual-level risk estimates may be considerable.

**Methods**

We propose a decomposition of Fisher's information matrix to help examine and calculate the sample size required for developing a model with acceptably precise risk estimates to inform decisions and improve fairness. We outline a five-step process for use before data collection or when an existing dataset is available. It requires researchers to specify the overall risk in the target population, the (anticipated) distribution of key predictors in the model, and an assumed 'core model' either specified directly (i.e., a logistic regression equation is provided) or based on a specified *C*-statistic and relative effects of (standardised) predictors.

**Results**

We produce closed-form solutions that decompose the variance of an individual's risk estimate into the Fisher's *unit* information matrix, predictor values and the total sample size; this allows researchers to quickly calculate and examine individual-level uncertainty interval widths for specified sample sizes, alongside prediction and classification instability plots. The information can be presented to key stakeholders (e.g., health professionals, patients, grant funders) to determine a target sample size or whether an existing dataset is sufficient. Our proposal is implemented in our new software module *pmstabilityss*. We provide two real examples and emphasise the importance of clinical context, including any risk thresholds for decision making, and fairness checks.

**Conclusions**

Our approach helps researcher identify the (target) sample size required to improve trust, reliability and fairness in individual-level predictions for binary outcomes.

## 1. Background

Studies developing a clinical prediction model use a sample of data from a chosen target population (e.g., pregnant women; men diagnosed with prostate cancer) to produce a model for predicting an outcome value (e.g., birth weight) or estimating an outcome risk (e.g., 30-day mortality risk) in any individual from that target population. Models are created using approaches such as regression, random forests or deep learning, which map predictor information to outcomes at the individual level. An example is the ISARIC model,[1] for use in hospitalised adults with suspected or confirmed COVID-19 to estimate their risk of in-hospital clinical deterioration based on 11 predictors measured at hospital admission.

When developing a prediction model there is a responsibility to implement rigorous standards in study design and analysis.[2-4] One important aspect is the sample size used to develop and evaluate these models. Many prediction model studies fail to include a justification of their sample size[5-12], despite being a recommendation in the TRIPOD statement,[13] and in the recently updated TRIPOD+AI guidance.[14] However, sample size criteria have been proposed in recent years for model development.[15-19] In previous work for binary outcomes,[16, 17] we outlined how to calculate the minimum sample size needed for model development based on (i) estimating the overall event risk precisely, and (ii) minimising model overfitting for a regression-based prediction model in terms of overall fit and population-level calibration slope. This criteria aims to target a more reliable prediction model at least at the *population level*, corresponding to the first two model stability levels defined by Riley and Collins.[20]

However, even when meeting this recommended minimum sample size, the uncertainty in a model's *individual-level* predictions can still be large. Individual-level predictions are a function of all the parameters (e.g. intercept, predictor effects) in the developed model, and large uncertainty of the parameter estimates leads to concerns of model instability,[20-22] with subsequently wide uncertainty intervals (e.g., 95% confidence or credible intervals) around individual estimated risks. Models that exhibit unacceptably high levels of uncertainty in their estimated risks should not be recommended for use in individuals, as point estimates of risk might misinform treatment and other healthcare decisions. Thus, further sample size calculations would be helpful to address precision of risk estimates from prediction models at the individual level, not just population level. Such calculations would also help improve *fairness* of prediction models,[23] such that the reliability (accuracy) of predictions is expected to be acceptable for all patient groups, including minoritised and underserved groups, not just in the population as a whole.[24]

In this article, we propose a decomposition of Fisher's information matrix to help researchers examine and estimate the sample size required to target sufficiently precise individual-level predictions, corresponding to the third and fourth stability levels of Riley and Collins.[20] Specifically, the decomposition leads to closed-form solutions in terms of Fisher's *unit* information matrix and total sample size, which can be used (either before data collection or when an existing dataset is available) to examine how sample size impacts the widths of uncertainty intervals around individual risk estimates from a model with the core predictors included. When a risk threshold is used to guide clinical decision making, the subsequent instability in classifications can also be examined for specified sample sizes.

The article outline is as follows. In Section 2 we summarise our existing sample size approach and the software module *pmsampsize* that implements it. Then, in Section 3 we outline our new proposal to examine sample size requirements to target precise and fair individual-level predictions. Section 4 applies our proposal to two real examples, one before data collection and one when an existing dataset is available. Section 5 concludes with discussion.

2. Existing sample size approach to precisely estimate overall risk and minimise overfitting for a binary outcome

Our current approach calculates the minimum required sample size for prediction model development,[15-17] to meet the following criteria:

- criterion (i): a precise estimate of the overall outcome risk
- criterion (ii): small overfitting of predictor effects
- criterion (iii): small optimism in apparent model fit

For brevity, details of the calculations are provided in supplementary material S1 and our previous papers.[15-17, 25] The approach is implemented in the Stata or R module *pmsampsize*,[26, 27] with the user needing to specify the overall outcome risk (prevalence) and the anticipated model performance (quantified by Cox-Snell R-squared ($R^2_{CS}$), Nagelkerke R-squared ($R^2_{Nagelkerke}$), or the *C*-statistic) in the target population, and the number of candidate predictor parameters for the model. For example, in Section 4.1 we consider development of a model to estimate the risk of foot ulcer based on three predictor parameters. Assuming a C-statistic of 0.77 and an overall outcome risk of 0.059 (based on previous studies[28]), then *pmsampsize* (available in Stata or R) calculates at least 453 participants (27 events) are needed to meet criteria (i) to (iii).

Also, in Section 4.2 we consider the development of a model to estimate the risk of acute kidney injury based on nine predictor parameters. Assuming an overall risk of 0.174 and a C-statistic of 0.78 based on a previous study,[29] then *pmsampsize* calculates that at least 511 participants (89 events) are required.

3. Methods: A new approach using a decomposition of Fisher's information matrix

Criteria (i) to (iii) of Section 2 aim for stable (and well-calibrated) model predictions at the population level,[20] such that the overall risk is precisely estimated and overfitting is low. As these are population-level criteria, they provide a *minimum* sample size calculation for model development. To extend calculations to the individual-level, we now propose a five-step process which utilises simulated (synthetic) or existing data, and a decomposition of the variance of individual-level predictions into Fisher's unit information matrix and total sample size. This five-step process is described below, and implemented using our new *pmstabilityss* module in Stata (type: net from https://joieensor.github.io/pm-suite/) or see (https://github.com/JoieEnsor). An R version is in development.

3.1 Five-step approach to calculating and examining individual-level prediction and classification uncertainty for specified sample sizes

**Step (1) - identify a core set of predictors:**

These core predictors are variables well-known (in the clinical setting of interest) to contribute important predictive information, for instance as identified from previously published models, systematic reviews of prognostic factors,[30] or conversations with clinical experts. Though additional predictors might be considered in the actual model development, the set of core predictors (and their combinations of values therein) represent the user's minimum for examining sample size and uncertainty of individual-level risks.

Examples of core predictors include (i) age and stage of disease for cancer outcome prediction; (ii) age, cholesterol and SBP for cardiovascular disease prediction; and (iii) insensitivity to monofilament, foot pulse absence, and history of previous ulcer or amputation for diabetic foot ulcer prediction.[28]

In addition, the core set may also include variables linked to fairness checks. For example, it may be important to ensure a developed model has sufficiently precise predictions for different demographic subgroups, such as age, sex and ethnicity or other protected characteristics. Ensuring these variables are included in the core predictor set can help to examine this, even if they are deemed to not be important predictors themselves.

**Step (2) - specify the joint distribution of the core predictors:**
The joint distribution of predictors that are included in the final prediction model impacts the standard errors of logistic regression parameter estimates, and thus influences the width of uncertainty intervals around individual-level risk estimates. Hence, step (2) requires the user to specify the joint distribution of core predictors selected in step (1). How to achieve this will depend on the availability of the model development dataset, as follows

- ***Dataset is already available for model developers***: this is a common situation and the most simple, as the joint distributions are already observed and so the user does not need to do anything in this step; further, the existing dataset can be used directly in subsequent steps where needed (e.g., in step (4) to derive the unit information matrix). See example of Section 4.2.
- ***Dataset exists but not yet available for model developers (e.g., access to a previously collected dataset is conditional on funding success):*** the data holders could be contacted and asked to provide a synthetic dataset that mimics the joint predictor distributions, for instance as obtained by a simulation-based approach, using packages such as *synthpop* in R.[31] A notable example is The Clinical Practice Research Datalink (CPRD), who have generated synthetic datasets to aid researchers improve workflows (https://www.cprd.com/synthetic-data). At the very least, data holders could provide summary details of the joint distributions (e.g., cross-tabulations of categorical variables; variance-covariance matrix of continuous variables), to allow the user themselves to simulate a large synthetic dataset containing predictor values for, say, 10000 individuals. See example in Section 4.1. Previous studies that already use the dataset of interest may summarise baseline variables (e.g., means, SDs, proportions) in their articles, for example within 'Table 1'.

- ***New data collection required (e.g., a planned prospective cohort study):*** summary information (e.g., means and SDs for continuous predictors; proportions in each group for categorical predictors) could be obtained from previous studies in the same target population, for example using published tables of baseline characteristics. These can then be used to simulate a large synthetic dataset of predictor values for, say, 10000 individuals. A challenge is that often only the marginal distribution for each predictor will be summarised (e.g., from a published 'Table 1' of baseline characteristics), and the correlation or conditional relationship amongst predictors will be unknown. In this situation, a pragmatic starting point is to assume the predictors are conditionally independent, but the impact of this should be examined (see example in Section 4.2).

**Step (3) - specify a 'core model' for how individual risks depend on core predictor values:** Alongside an individual's predictor values, the uncertainty around an individual's risk estimate also depends on their risk estimate itself. Therefore, step (3) requires the user to specify a model that expresses how an individual's risk depends on the values of core predictors from step (2). We refer to this as the 'core model'. For example, one could specify a logistic regression model, such that an individual's logit-risk is a function defined by an intercept and beta terms chosen to reflect the overall population risk and core predictor effects, respectively. This could be based on a previous model in the same field, and Section 4.1 gives such an example. This is especially relevant when updating or extending an existing model.

Often a 'core model' may be difficult to specify from scratch; for example, predictor effects may not be readily available if previous models were unreported or 'black box'. To address this, we now outline two approaches to simplify the process, that utilise the anticipated C-statistic (equivalent to the area under the receiver operating characteristic curve, AUROC) of the model and overall risk in the target population, alongside assumptions of *relative* predictor effects (weights).

- *Approach (a): Specify the overall risk, C-statistic, and relative weights of core predictors.* With this information, the approach based on Austin[32] can be used to identify a logistic regression equation that forms a 'core model' that adheres to the specified overall risk and C-statistic, whilst retaining the user's chosen *relative* weight of core predictors. In brief, the approach simulates predictor values for a large number of participants (based on the distributions specified in Step (2)) and then an iterative process is used to identify values of the intercept ($\alpha$) and a multiplicative factor ($\delta$) of the following model,

$$y_i \sim \text{Bernoulli}(p_i)$$
$$\ln\left(\frac{p_i}{1-p_i}\right) = \alpha + \delta(\beta_1 x_{1i} + \beta_2 x_{2i} + \cdots + \beta_P x_{Pi}) \qquad \text{Eq. (1)}$$

  where the beta coefficients are the relative weights specified by the user. Convergence is achieved when the model has reached the specified *C*-statistic and overall risk within a small margin of error.

- *Approach (b): specify the overall risk and C-statistic, whilst assuming same weight of predictors after standardising continuous predictors.* This approach is akin to approach (a) but simplifies the process by assuming the 'core model' of Eq. (1) has equal weight of all predictors (i.e. all betas are set to 1) after standardising continuous predictors using their marginal mean and standard deviation from the joint distribution specified in Step (2) (e.g., uses $(x_{1i} - \bar{x}_{1i})/\text{SD}_{x_{1i}}$ when $x_{1i}$ is a continuous variable, etc.). As in Approach (a), the Austin method is then used to identify $\alpha$ and $\delta$ that ensures the 'core model' has a particular *C* statistic and overall risk. An example is shown in Section 4.2 for a model with nine predictors and the approach is particularly relevant when there is little information about predictor weights and relative importance of predictors from previous studies. Categorical variables could also be standardised, but we generally prefer to include them on their original scale by default, so that the predictive effect of one category (e.g., males) relative to the reference category (e.g., females) is equivalent to a 1-SD increase in the continuous variable.

**Step (4) - derive Fisher's unit information after decomposing Fisher's information matrix:**

Steps (4) and (5) involve approximating (based on the information from steps (1) to (3)) the anticipated variance-covariance matrix (var($\widehat{\boldsymbol{\beta}}$)) of model parameter estimates ($\widehat{\boldsymbol{\beta}} = (\hat{\alpha}, \hat{\beta}_1, \hat{\beta}_2, \ldots, \hat{\beta}_P)'$) if we were to fit the assumed 'core model' for a specified sample size. This is needed, as the variance-covariance matrix subsequently dictates the variance of individual-level predictions. Step (4) begins this process by decomposing var($\widehat{\boldsymbol{\beta}}$) (i.e., the inverse of Fisher's information matrix) into the total sample size ($n$) and Fisher's <u>unit</u> information matrix (**I**),

$$\text{var}(\widehat{\boldsymbol{\beta}}) = n^{-1}\mathbf{I}^{-1} \qquad \text{Eq. (2)}$$

where,

$$\mathbf{I} = E\left(\frac{\exp(\mathbf{X}'\boldsymbol{\beta})}{(1+\exp(\mathbf{X}'\boldsymbol{\beta}))^2}\mathbf{X}\mathbf{X}'\right) = E(\mathbf{A}) \qquad \text{Eq. (3)}$$

and **X** is the design matrix for the assumed 'core model', with each individual's data corresponding to $\boldsymbol{x_i} = (1, x_{1i}, x_{2i}, \ldots, x_{Pi})$.

The expected value ($E(\mathbf{A})$) depends on the joint distribution of the predictors and the parameter values of the 'core model'. A simple way to derive $E(.)$ is to calculate each of the components of **A** for each participant in the (existing or simulated) dataset from Step (2), using each participant's predictor values combined with the logistic regression parameters from Step (3); then the means (across all participants) of each component provides their expected values and forms **I**.

For example, assuming a model with three core predictors we can write,

$$\begin{aligned}\mathbf{I} &= E\left(\frac{\exp(\alpha + \delta(\beta_1 x_{1i} + \beta_2 x_{2i} + \beta_3 x_{3i}))}{(1+\exp(\alpha + \delta(\beta_1 x_{1i} + \beta_2 x_{2i} + \beta_3 x_{3i})))^2}\mathbf{X}\mathbf{X}'\right) \\ &= E\left(\frac{\exp(\alpha + \delta(\beta_1 x_{1i} + \beta_2 x_{2i} + \beta_3 x_{3i}))}{(1+\exp(\alpha + \delta(\beta_1 x_{1i} + \beta_2 x_{2i} + \beta_3 x_{3i})))^2}\begin{bmatrix} 1 & x_{1i} & x_{2i} & x_{3i} \\ x_{1i} & x_{1i}^2 & x_{1i}x_{2i} & x_{1i}x_{3i} \\ x_{2i} & x_{1i}x_{2i} & x_{2i}^2 & x_{2i}x_{3i} \\ x_{3i} & x_{1i}x_{3i} & x_{2i}x_{3i} & x_{3i}^2 \end{bmatrix}\right) \qquad \text{Eq.(4)} \\ &= E(\mathbf{A})\end{aligned}$$

where **A** is a 4 by 4 matrix. Crucially, the parameters (i.e., $\alpha$, $\delta$, and all $\beta$s) are replaced with their true values specified in Step (3), and the $x$ values are the (standardised or unstandardised) values of core predictors from Step (2). Then, to derive $E(\mathbf{A})$ our software package calculates each of the 16 components of **A** for each participant in the existing or synthetic dataset, and the mean of each component (over all participants) provides their expected values and thus forms **I**.

**Step (5) - examine the impact of sample size on precision of individual risk estimates:**
The final step is to examine how sample size impacts the level of precision (uncertainty interval widths) around individual risk estimates. This is relevant when the user has an existing dataset (e.g., to ascertain if it is large enough) or when designing a new study with prospective data collection (to *a priori* identify the required sample size). These situations are now outlined as Options A and B, below.

- *Option A: Calculate expected uncertainty of predictions for a given sample size (existing dataset)*

Following maximum likelihood theory of a logistic regression model, the variance of a predicted logit-risk for a new individual is,

$$\text{var}(\text{logit}(\hat{p}_{new})) = \text{var}(\boldsymbol{x}_{new}\widehat{\boldsymbol{\beta}}) = \boldsymbol{x}_{new}\,\text{var}(\widehat{\boldsymbol{\beta}})\,\boldsymbol{x}'_{new} \qquad \text{Eq. (5)}$$

where $\boldsymbol{x}_{new} = (1, x_{1new}, x_{2new}, \dots, x_{Pnew})$ are the predictor values for the new individual. Substituting in Eq. (2), this can be rewritten as:

$$\text{var}(\text{logit}(\hat{p}_{new})) = n^{-1}\,\boldsymbol{x}_{new}\,\mathbf{I}^{-1}\,\boldsymbol{x}'_{new} \qquad \text{Eq. (6)}$$

Subsequently, a 95% uncertainty interval around an individual's estimated risk is,

$$\text{invlogit}\left[\text{logit}(\hat{p}_{new}) \pm \left(1.96 \times \sqrt{\text{var}(\text{logit}(\hat{p}_{new}))}\right)\right] \qquad \text{Eq. (7)}$$

and substituting in Eq. (6) gives,

$$\text{invlogit}\left[\text{logit}(\hat{p}_{new}) \pm \left(1.96 \times \sqrt{(n^{-1}\,\boldsymbol{x}_{new}\,\mathbf{I}^{-1}\,\boldsymbol{x}'_{new})}\right)\right] \qquad \text{Eq. (8)}$$

where 'invlogit' is the inverse of the logit function (i.e., $\exp(\hat{p}_{new})/(1+\exp(\hat{p}_{new}))$).

Option A requires the user to apply Eq. (8) to participants from the target population, to derive uncertainty intervals conditional on a particular model development sample size ($n$) of interest. We already have the unit information matrix (**I**) from Step (4), and the new participants can just be those from the existing or simulated dataset from Step (2) which already contain predictor values ($\boldsymbol{x}_{new}$). Each individual's $\hat{p}_{new}$ can be set to their 'true' risk defined by the 'core model' in Step (3). Thus, to apply Eq. (8) we just need to specify the sample size ($n$) of interest: this could be the available number of participants in an existing dataset being considered for model development, or it might be a specified sample size being considered for new data collection (e.g., determined by *pmsampsize* for criteria (i) to (iii)). Sometimes, a range of different sample sizes might be considered to examine the value of information (e.g., in terms of reduced width of uncertainty intervals, reduced classification instability) arising from including additional participants over and above that recommended by *pmsampsize*.

- ***Option B: Calculate a target sample size for new data collection to ensure precise individual-level predictions***

When designing a new study to recruit participants for model development, researchers will want to calculate the sample size required to target particular precision of risk estimates. By rearranging Eq. (6), the sample size needed to target a chosen variance of the logit-risk estimate for an individual is:

$$n = \text{var}(\text{logit}(p_{new}))^{-1} \boldsymbol{x}_{new} \ \mathbf{I}^{-1} \ \boldsymbol{x}'_{new} \qquad \text{Eq. (9)}$$

We can apply this to every individual in the (real or simulated) dataset from Step (2) to obtain the required $n$ for their particular combination of predictor values ($\boldsymbol{x}_{new}$). A practical issue is how to select the target value of $\text{var}(\text{logit}(p_{new}))$ for each individual, as this is on a difficult scale to interpret. Further, the required value of $\text{var}(\text{logit}(p_{new}))$ will not be consistent across individuals, due to the multiplication with $\boldsymbol{x}_{new} \ \mathbf{I}^{-1} \ \boldsymbol{x}'_{new}$, which is individual specific. A pragmatic approach is to specify the maximum $\text{var}(\text{logit}(p_{new}))$ allowed for a range of $p_{new}$ values (e.g., 0.01, 0.025, 0.05, 0.10, 0.15, 0.20, etc), corresponding to a target maximum uncertainty interval width on the risk scale (via Eq. (8)).

Eq. (9) can then be applied to each individual by using the $\text{var}(\text{logit}(p_{new}))$ value that corresponds to the categorised $p_{new}$ value closest to their estimated $p_{new}$. An example of this process is provided in Section 4.1. Special attention may also be given to selecting appropriate $\text{var}(\text{logit}(p_{new}))$ values in particular subgroups defined by combinations of predictor values (e.g., sex, ethnicity), where algorithmic fairness checks will be important.

3.2 Deciding and presenting target uncertainty intervals with patients and clinical stakeholders: perspective based on risk thresholds and decision analysis theory

Regardless of whether option A or B is chosen, model developers will need to decide what width of uncertainty intervals they deem appropriate. This is always context dependent for the clinical setting of interest; for example, it can depend on patient preferences about potential outcomes, treatments and consequences, and corresponding risk thresholds to inform decision making. This could be discussed with stakeholders (e.g., patients, doctors, health professionals and regulators) in advance of data analysis, to help identify what levels of uncertainty would lead to them recommending further research is still needed.[33]

Ideally, a suitably narrow interval is desired for *every* individual (for all combinations of predictor values). However, depending on the clinical context and role of the model for clinical practice, some regions of estimated risk may not require intervals to be as narrow as in other regions. For instance, having wide uncertainty intervals for individuals with high risk (e.g., reflected by uncertainty intervals from 0.3 to 0.95) may not matter if the entire interval range is still compatible with a perceived high risk. This concept aligns with preferences for risk thresholds for clinical decisions. For example, 10-year CVD risk thresholds of 10% are sometimes used to guide decisions to prescribe statins, and so wide uncertainty intervals that span 0.3 to 0.95 might be deemed acceptable, but narrower intervals that span 0.05 to 0.3 may not. Therefore, when deciding upon appropriate uncertainty interval widths, it is imperative to understand the clinical context of how the model will be used to guide decision making and any corresponding risk threshold(s) involved.

With this in mind, it is helpful to identify risk thresholds within a decision-theory perspective,[34-36] based on preferences (utilities) elicited from patients, clinicians and other relevant stakeholders about particular outcomes and consequences that may follow from possible decisions. An example of this process is given in supplementary material S2, where a prostate cancer risk threshold of 4.9% is identified for deciding when a particular individual warrants a biopsy. If a well-calibrated prediction model estimates the individual's risk to be > 4.9%, then this suggests the correct decision is to biopsy. However, there may still be uncertainty about this decision due to uncertainty of the model's risk estimate; if the uncertainty is too wide, then ideally further information is needed before making a decision. Vickers et al. support this argument,[33] noting that "decision analysis tells us which decision to make for now, but we may also want to know how much confidence we should have in that decision. If we are insufficiently confident that we are right, further research is warranted."

In this context, the aim of our sample size approach is to help understand and examine which sample sizes are likely to give sufficient information to guide decisions at the individual-level. To help examine this, we recommend calculating and presenting to stakeholders:

- *prediction instability plots*, where each individual's 'true' risk from Step 3 (x-axis) is plotted against their corresponding uncertainty interval (y-axis) from Step 5. The question to ask stakeholders is whether the individual uncertainty intervals are too wide for using or endorsing the model in practice. To facilitate this discussion, we recommend prediction instability plots are presented with two curves (e.g., using a LOWESS smoother or spline function) fitted separately through individuals' upper and lower uncertainty interval values. These curves define a 'typical' 95% uncertainty interval at each risk, across the entire spectrum of estimated risks from 0 to 1, which should aid visual interpretation for stakeholders (as individual uncertainty intervals can vary considerably, even for those with the same estimated risk). This is demonstrated in Section 4.2.
- *classification instability plots (if risk thresholds are relevant)*, plotting each individual's 'true' risk (x-axis) (i.e., that risk defined by the 'core model') against the proportion (y-axis) of their uncertainty distribution that falls on the opposite side of their chosen clinical risk threshold compared to their 'true' risk. The question to ask stakeholders is whether, in general, the proportion of the uncertainty distribution in the opposite direction is generally too large for them to use or endorse the model for individuals in practice. In our examples in Section 4, we assume the same risk threshold is relevant for all individuals, but this can be relaxed if stakeholders recommend different thresholds across particular subgroups.

- *summary statistics,* that quantify the magnitude of uncertainty and classification instability across individuals. In particular, the mean (min, max, median etc) width of 95% uncertainty intervals, and the mean (min, max, median etc) probability of misclassification. Also, the mean (min, max, median etc) across individuals of their mean absolute prediction error (MAPE), which can be derived by many (e.g., 1000) sampling values from each individual's uncertainty distribution and calculating mean absolute differences to their 'true' risk.
- *subgroup plots and results,* that summarise the anticipated uncertainty and classification instability in relevant subgroups of people (e.g., defined by relevant attributes including sex, ethnicity – see Section 5).

The value of perfect information is to remove any prediction uncertainty and thus alleviate the potential for any misclassification (compared to the 'true' risks based on the 'core model'). So, we need to ascertain how much uncertainty and potential misclassification (if relevant) is acceptable, and then identify the sample size required to achieve this (or whether an existing dataset achieves this). We now illustrate this idea with two examples applying our new approach.

4. Results I: Applied examples

We now consider two examples; the first considers a situation in advance of data collection and the second considers when an existing dataset is available. The examples can be implemented using a single line of code via our *pmstabilityss* module in Stata (https://github.com/JoieEnsor).

4.1 Example 1: Prediction model for foot ulcer

Chappell et al developed a prediction model for risk of foot ulcer by 2 years in people with diabetes,[28] which contained three predictors. Let us assume that a new cohort study is planned to collect new data to update and potentially extend this model. What sample size should be targeted? Researchers should start by considering criteria (i) to (iii) (to target population-level stability), for which *pmsampsize* suggests a minimum sample size of 453 participants (27 events), based on an overall risk of 0.059 and three predictor parameters (see Section 2). However, we should also examine the expected uncertainty of individual-level predictions and whether a larger sample size is needed for more precise and stable risk estimates. To examine this, we applied the five-step process of Section 3, as follows.

*Step 1: Identify core predictors*

The core predictors were chosen to be all those included in Chappell's model: mono (1 if insensitive to monofilament), pulse (1 if missing at least one foot pulse) and history (1 if previous ulcer or amputation). Although other predictors might be considered in the new model, it was deemed fundamental to examine individual-level uncertainty of risk estimates across combinations of these three predictors.

*Step 2: Specify joint distribution of the predictors*

Upon request, Chappell et al. summarised the joint distribution of these three binary predictors in their model development dataset:

- Mono 0 Pulse 0 History 0 (56.3%)
- Mono 1 Pulse 1 History 1 (2.1%)
- Mono 1 Pulse 1 History 0 (6.2%)
- Mono 1 Pulse 0 History 1 (3.2%)
- Mono 1 Pulse 0 History 0 (11.5%)
- Mono 0 Pulse 1 History 1 (1.2%)
- Mono 0 Pulse 1 History 0 (17.7%)
- Mono 0 Pulse 0 History 1 (2.0%)

We simulated a dataset of 10,000 participants with predictor values randomly sampled from this joint distribution.

*Step 3: Specify the 'core model'*

The 'core model' was assumed to be the same as Chappell et al:

$$\ln\left(\frac{p_i}{1-p_i}\right) = 3.81 + (1.11 \times mono) + (0.70 \times pulse) + (1.95 \times history)$$

For every individual in the simulated dataset, we used this model to generate their true (logit) risk conditional on their simulated predictor values from Step 2.

***Step 4: Derive Fisher's unit information after decomposing Fisher's information matrix***

The unit information matrix was calculated using Eq. (3) (equivalent to Eq.(4) for this three predictor model), by calculating each of the components of **A** for each individual in the simulated dataset from step (2) (conditional on their predictor values and 'true' risks), and then taking the means (across all participants) of each component to give expected values and thus form **I**. This is embedded in our software module (see below).

***Step 5: Identify the minimum sample size needed to achieve narrow uncertainty intervals***

Finally, options A and B of Step 5 were both applied to examine and determine the sample size required for appropriately narrow uncertainty intervals at the individual level. This required multiple investigations, with consideration of clinical context, as follows.

*(i) Applying Option A*

Option A can be used to calculate anticipated uncertainty intervals of individual risk based on a user-specified sample size, which is implemented in our accompanying *pmstabilityss* module. It requires the (simulated) dataset from Step 2; the variable names denoting the core predictors (Step 1); and the overall risk, sample size and assumed betas of the 'core model' (Step 3). It then implements Step 4 and Option A of Step 5, to derive uncertainty intervals for all individuals in the dataset, summary statistics about the uncertainty intervals and MAPE (e.g., mean, minimum and maximum), and prediction and classification instability plots. The code is provided elsewhere (https://github.com/JoieEnsor).

In this example, there are 8 groups of participants (defined by the 8 combinations of values of the 3 binary predictors). Assuming a total sample size of 453 participants (the minimum recommended based on criteria (i) to (iii)), anticipated 95% uncertainty intervals for each group are shown by the prediction instability plot of Figure 1(a). The mean (min, max) interval width is 0.08 (0.03, 0.42) with, as expected, wider intervals for those with less common combinations of predictor values. In particular, individuals with a 'true' risk of 0.49 have the least common set of predictor values (mono 1, plus 1, history 1), leading to a wide 95% uncertainty interval of 0.28 to 0.70. The mean (min, max) MAPE is 0.017 (0.0003, 0.20).

Alongside individual-level uncertainty, we can also examine instability in subgroups defined by one particular predictor value (essentially integrated over the distribution of the other two predictors in the model). For example, in those missing at least one foot pulse, the average uncertainty interval width is 0.14 compared to 0.06 for those with no missing foot pulse; and the mean MAPE is 0.027 compared to 0.013. Thus predictions are more uncertain for those missing a foot pulse. Subgroups are discussed more in Section 5.

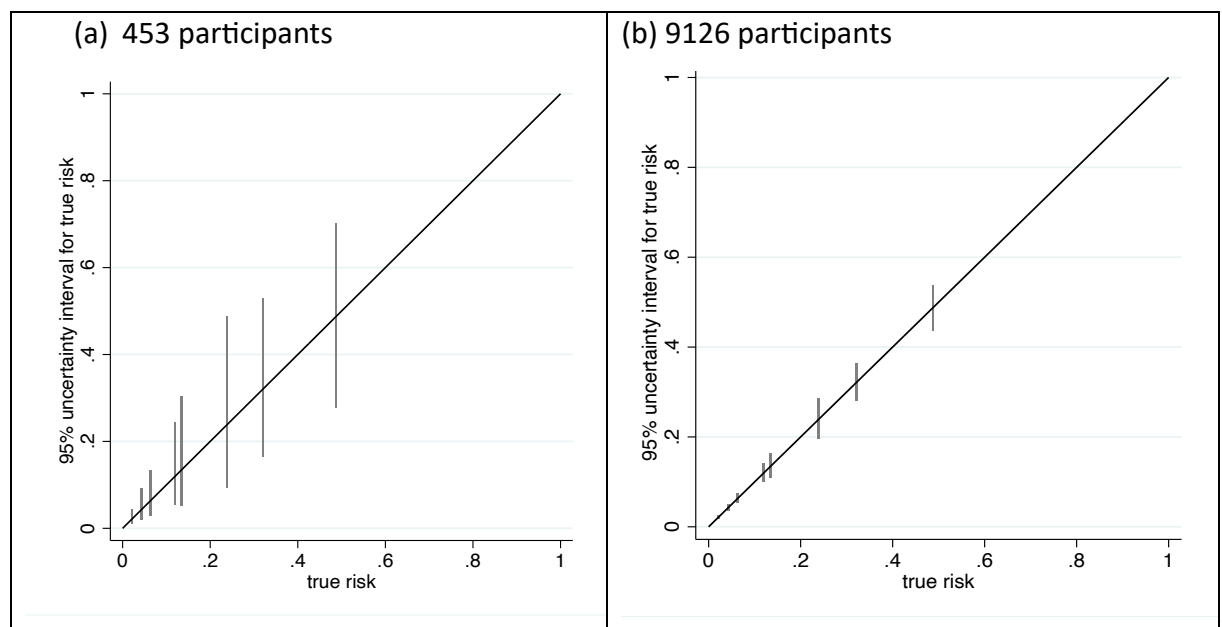

*Figure 1 Anticipated uncertainty interval widths when developing a foot ulcer prediction model with a sample size of (a) 453 participants and (b) 9126 participants, based on the prediction model of Chappell et al.*

### *(ii) Applying Option B to target uncertainty interval widths of 0.1 or less*

Now consider using Option B to calculate the sample size required to target uncertainty interval widths $\leq 0.1$ for *all* individuals. This can be calculated by applying Eq. (9) and is implemented in *pmstabilityss* (https://github.com/JoieEnsor). This calculates that a sample size of 9126 participants is required, which is driven by the group with a risk of 0.49 who, as mentioned, have the least frequent set of predictor values. The corresponding prediction instability plot for 9126 participants is shown in Figure 1(b) and, as targeted, all interval widths are 0.1 or less.

### *Decision-analysis perspective: identifying the target sample size considering clinical context and classification instability*

Although the Option B sample size of 9126 participants would be ideal to target narrow intervals for everyone, it may not be achievable or even necessary. Consideration of clinical context and decision thresholds is key here. In their paper, Chappell et al suggest a risk threshold of about 6% for deciding when individuals should be prescribed preventative treatment for foot ulcers. Therefore, an interval width of 0.1 for everyone may be too stringent, especially those individuals with a predicted risk of 0.49, as the 6% threshold is far from their actual risk. Even if their uncertainty interval width was, say, 0.2 or 0.3, then the interval would still be precise relative to the threshold.

We can examine this by using a classification instability plot, which *pmstabilityss* produces if the user specifies the threshold value, which here is 6%. Figures 2(a) and 2(b) show classification instability plots for the two sample sizes of 453 and 9126 participants, respectively. They reveal the proportion of each individual's uncertainty distribution that is on the other side of the threshold compared to their 'true' risk (i.e., their probability of misclassification). For individuals with a 'true' risk of 0.063 close to the threshold, a high proportion of their uncertainty distribution is on the opposite side of the threshold to their 'true' risk. Such uncertainty close to the threshold is inevitable unless extremely large sample sizes are used. However, for other individuals, the classification instability is low even in the minimum sample size of 453 participants recommended by *pmsampsize*. For example, for those with a 'true' risk of 0.13, only about 5% of their uncertainty distribution is below the threshold when the sample size is 453 participants, and close to 0% with 9126 participants. For either sample size, the instability is close to 0% for individuals with 'true' risks close to 0 or above 0.2. Therefore, the sample size of 453 participants (identified from *pmsampsize* based on criteria (i) to (iii)) still seems appropriate to obtain narrow-enough uncertainty intervals in the context of a risk threshold of 6%. The value of additional information beyond this appears slight.

However, what if stakeholders had suggested a threshold of 20% was relevant to consider? For example, it might be that available treatments are very expensive or have side effects that warrant a higher risk threshold for action. This higher threshold may lead to a larger required sample size, because with 453 participants half the individuals still have uncertainty intervals that considerably overlap the 20% threshold (Figure 1(a)), and so the classification instability index is higher than before (Figure 3(a)). In particular, we might be concerned about the uncertainty interval width for the individuals with a 'true' risk of 0.13 (the mono = 0, pulse = 0, history = 1 group), as their classification instability index is large.

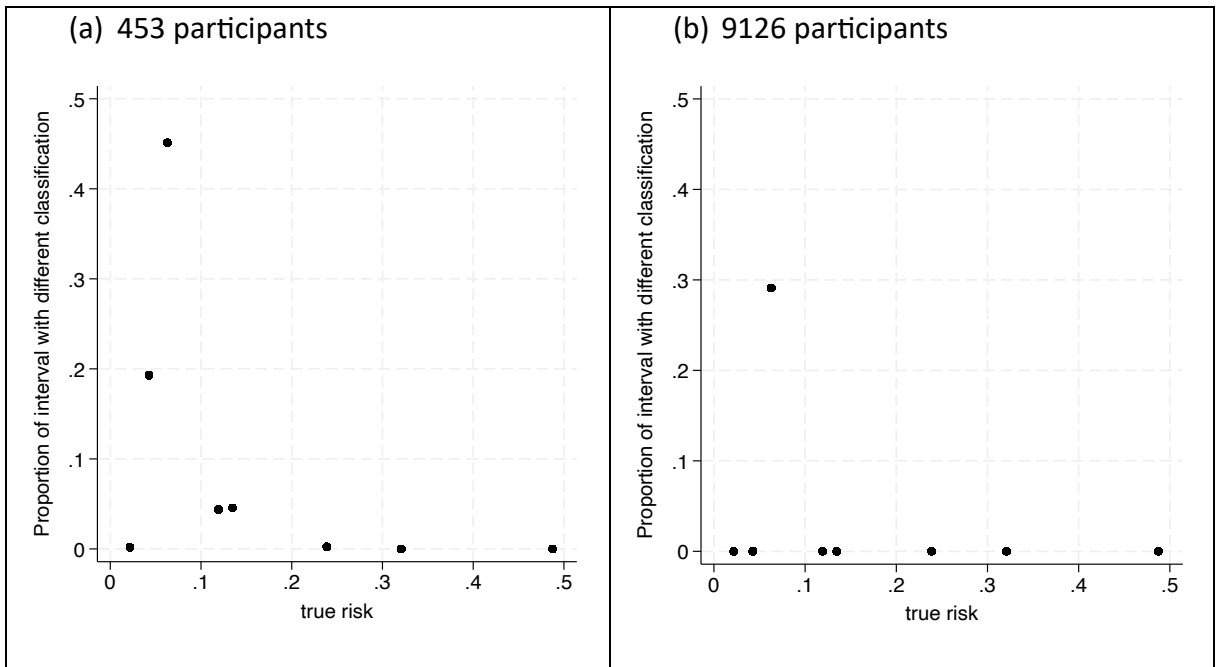

*Figure 2 Classification instability plots based on developing the foot ulcer prediction model for (a) 453 participants and (b) 9126 participants, showing the proportion (y-axis) of each individual's uncertainty distribution that is on the opposite side of the 6% risk threshold compared to their 'true' risk (x-axis)*

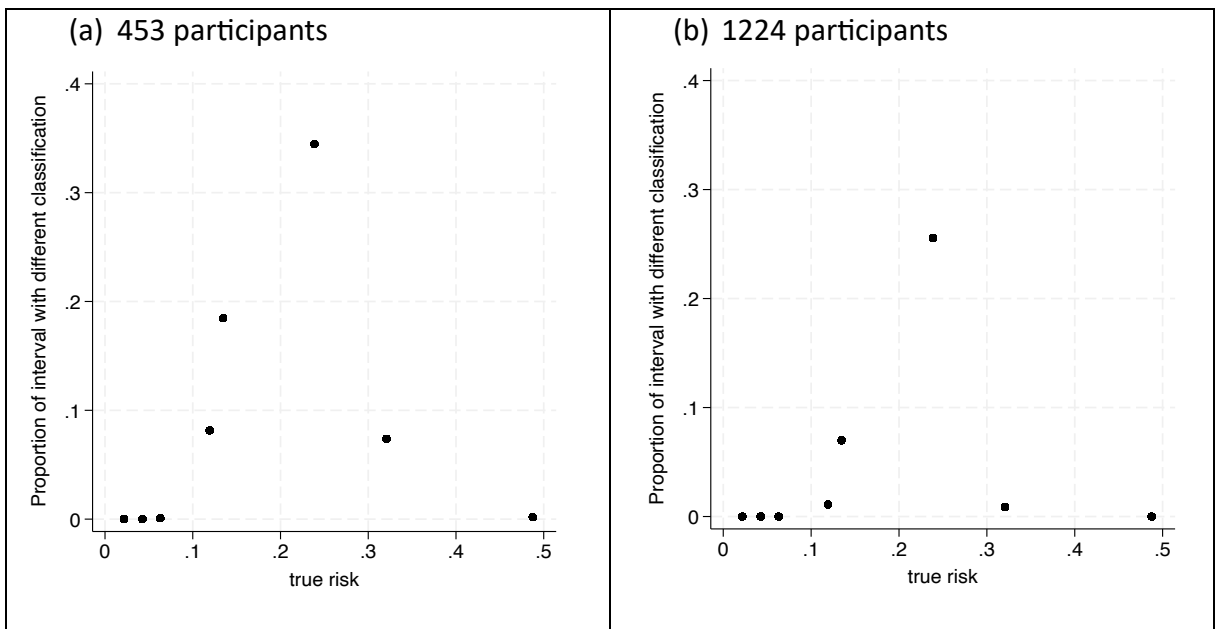

*Figure 3 Classification instability plots when developing the foot ulcer prediction model with (a) 453 participants and (b) 1224 participants, showing the proportion (y-axis) of each individual's predictive distribution that is on the opposite side of the 20% risk threshold compared to their 'true' risk (x-axis)*

To address this, Option B can be used to target a narrower uncertainty interval width of, say, 0.15 (a target $\text{var}(\text{logit}(p_{\text{new}}))$ of about 0.103) for this group of people who have a 'true' risk of 0.13, and applying Eq. (9) leads to a required sample size of 1224 participants. The classification instability index is now much lower (Figure 3(b)). Hence, if the threshold of 20% was of interest, a final sample size of 1224 participants might seem sensible, as it targets narrow uncertainty intervals and low classification instability for most individuals.

4.2 Example 2: Prediction of acute kidney injury from an existing dataset

Now, we present an example where an existing dataset (representative of the target population) is already available to researchers who want to examine (and perhaps justify to grant funders) its suitability for developing a model for use in intensive care patients, to estimate an individual's risk of acute kidney injury within 48 hours. The dataset was obtained from the Medical Information Mart for Intensive Care III (MIMIC-III),[37] which contains freely available and de-identified critical care data from the Beth Israel Deaconess Medical Center in Boston, Massachusetts, between 2001 and 2012. A cohort of size 20413 patients was extracted from this database, containing individuals aged over 18 years of age who were admitted to the intensive care unit for any cause for at least 24 hours. Acute kidney injury (the outcome of interest) was defined as present if the maximum creatinine within 48 hours after the prediction time (end of first day on ICU) was either: (i) more than 1.5 times the minimum day 1 creatinine value, or (ii) over 0.3 mg/dL greater than the minimum day 1 creatinine value.[38] The overall risk of acute kidney injury was 17% in the dataset.

The researchers want to ascertain (in advance of any analysis) whether this existing dataset is likely to produce a model with suitably narrow uncertainty intervals around individuals' estimated risks, and classifications based on estimated risk. The five-step process from Section 3 can be used to do this, as follows.

***Step 1: Identify core predictors***

There is a general lack of consensus on the key predictors in this field,[39] but the existing dataset contained nine variables included in a previous model by Zimmerman et al.[29], so these were selected as the core predictor set (age, sex, bicarbonate, creatinine, haemoglobin, blood urea nitrogen, potassium, systolic blood pressure, and SpO2).

***Step 2: Specify joint distribution of the predictors***

As the existing dataset was available, the joint distribution of these 9 predictors could be observed directly, and thus no simulated data was necessary. We can simply use the existing dataset at hand.

*Step 3: Specify the 'core model'*

Previous models in this field have been quite poorly reported, with very little information on sizes of predictor effects or the scale that predictors are measured on. In this situation, a pragmatic approach is to set all core predictors in the 'core model' to have the same weight, after standardising continuous predictors (see Approach (b) of Step 3 in Section 3.1), and to set the 'core model' to have a particular C-statistic and overall risk. Further, the *direction* of effect was based on the effect sizes reported by Zimmerman et al.[29], and all continuous predictors were assumed to have a linear effect for simplicity. Thus, the 'core model' was defined by,

$$\ln\left(\frac{p_i}{1-p_i}\right) = \alpha + \delta(1.\,\text{age}_{1i} - 1.\,\text{male}_{2i} - 1.\,\text{bicarbonate}_{3i} + 1.\,\text{creatinine}_{1i} - 1.\,\text{haemoglobin}_{2i}$$
$$- 1.\,\text{nitrogen}_{3i} - 1.\,\text{potassium}_{1i} - 1.\,\text{SBP}_{2i} - 1.\,\text{SpO2}_{3i})$$

where each of the eight continuous predictors are standardised, so that their relative weights are the same (e.g., a 1-SD increase in SBP has the same weight as a 1-SD increase in age), and have the same weight as being male (compared to female). Then, our *pmstabilityss* module implemented the iterative process of Austin,[32] to identify $\alpha = -1.97$ and $\delta = 0.40$ as the required values to ensure the model corresponds to an overall risk of 0.174 and a C-statistic of 0.78 (based on that reported for the previous model by Zimmerman et al.[29]). Using this 'true' model, we calculated true (logit) risks for each of the 20413 individuals, conditional on their observed predictor values.

*Step 4: Derive Fisher's unit information after decomposing Fisher's information matrix*

We calculated the unit information matrix (**I**) using Eq. (3), by calculating each component of **A** for each participant in the existing dataset (using their observed predictor values combined with the 'true' risks from step (3)) and then taking the means (across all participants) of each component to form the entries of **I**.

*Step 5: Derive anticipated uncertainty intervals and classification instability*

Finally, Option A was used to calculate anticipated uncertainty intervals for each of the 20413 individuals in the existing dataset. Our *pmstabilityss* module calculates this. The user supplying the existing dataset from Step 2, and the coded variable names representing the core predictors from Step 1 and the 'core model' from Step 3; then *pmstabilityss* implements Steps 4 and 5 to produce prediction and classification instability plots (see https://github.com/JoieEnsor).

The prediction instability plot is shown in Figure 4, displaying all the individual uncertainty intervals and also smooth curves fitted through the lower and upper values of the intervals, to generate 'typical' 95% uncertainty intervals for each 'true' risk. Regardless of the 'true' risk, the uncertainty intervals are all very narrow, which should be reassuring to the researchers and potential funders. The mean (min, median, max) uncertainty interval width is 0.030 (0.00002, 0.026, 0.16) and MAPE is 0.0062 (0.000004, 0.0053, 0.033).

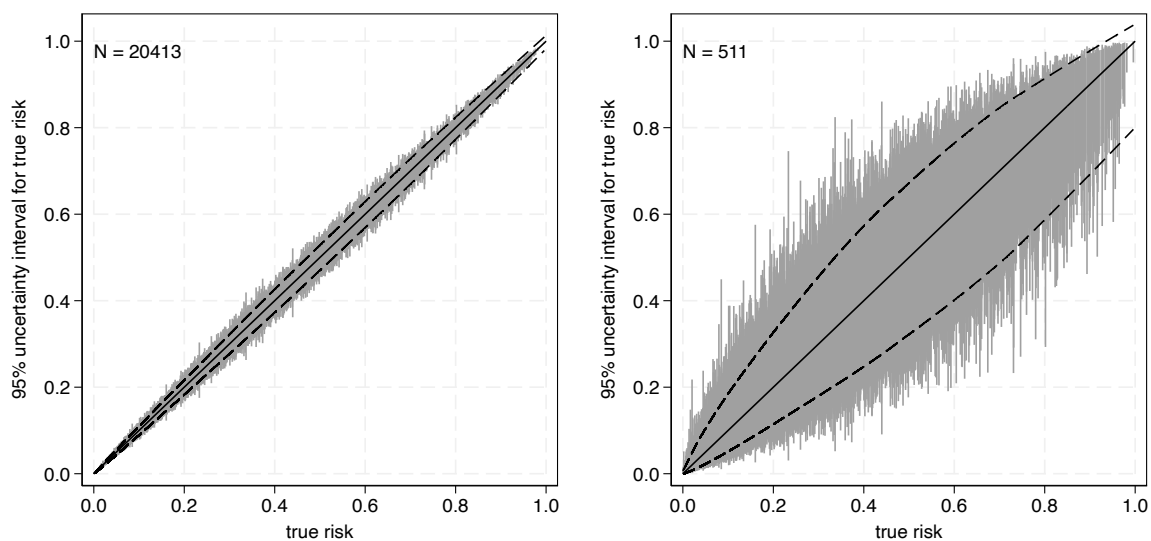

*Figure 4 Expected uncertainty interval widths when developing an acute kidney injury prediction model with a sample size of (a) 20413 participants (size of existing dataset) and (b) 511 participants (minimum size recommended by criteria (i) to (ii))*

### *Decision-analysis perspective*

Consider that stakeholders suggest a risk threshold of about 10% is relevant for informing clinical decisions, such as closer monitoring of urine output and renal function. The corresponding classification instability plot is shown in Figure 5, and the instability is close to zero for most individuals, except those very close to the assumed threshold of 10%. The mean (min, median, max) probability of misclassification is 0.013 (0, 0, 0.50). Thus, the existing sample size of 20413 is more than sufficient.

### *Comparison to pmsampsize recommended sample size of 511 participants*

For comparison, let us consider the prediction and classification instability plots had the existing dataset matched the sample size of criteria (i) to (iii), which *pmsampsize* calculates to be 511 participants. This is substantially smaller than the 20413 actually available, and leads to wide anticipated uncertainty intervals (Figure 4(a)) with larger classification instability (Figure 5(a)). The mean (min, median, max) width of uncertainty intervals is 0.19 (0.0047, 0.17, 0.77); probability of misclassification is 0.083 (0, 0.0067 0.50); and MAPE is 0.039 (0.0006, 0.034, 0.18). 'Typical' uncertainty intervals are depicted between the dashed lines; for example, someone who has a 'true' risk of 0.2 has a 'typical' uncertainty interval from about 0.1 to 0.4. Such wide intervals might reduce model acceptability and usefulness, and reinforces that the minimum sample size based on criteria (i) to (iii) does not necessarily guarantee narrow uncertainty intervals at the individual level. Nevertheless, discussions with stakeholders would be needed to properly ascertain whether the magnitude of uncertainty and classification instability is unacceptable, had the existing sample size been 511 participants.

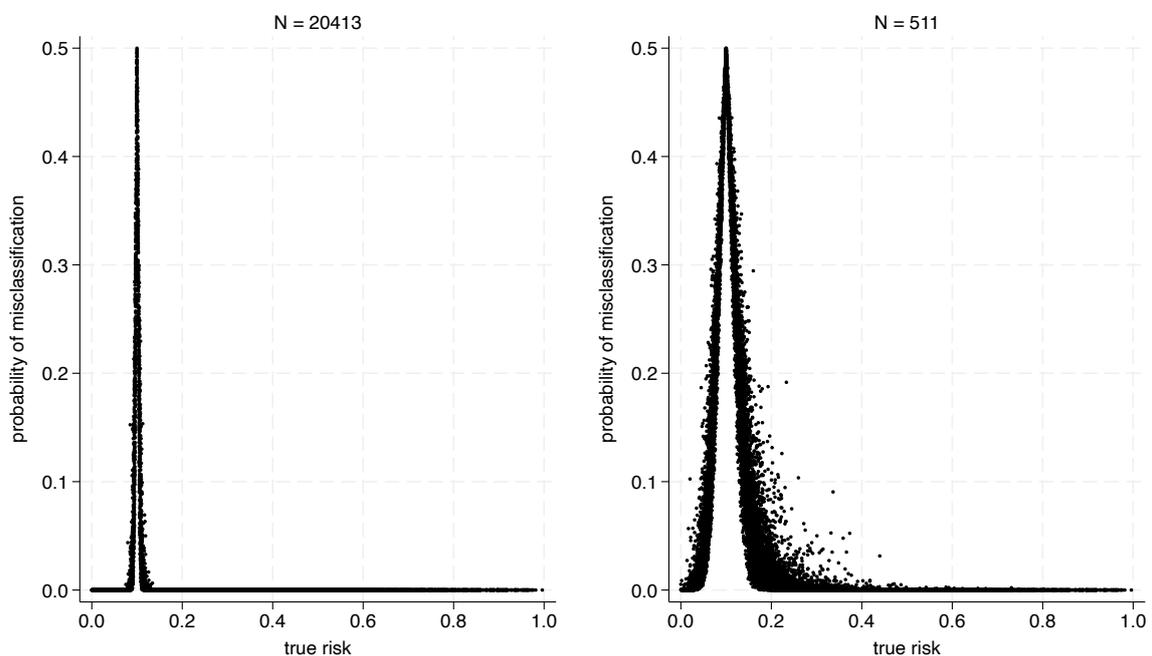

*Figure 5 Classification instability plots for the acute kidney injury model when developing a model with (a) (b) 20413 participants (size of existing dataset), or (b) 511 participants (minimum size recommended by criteria (i) to (ii)); the plot shows the proportion (y-axis) of each individual's uncertainty distribution that is on the opposite side of the 10% risk threshold compared to their 'true' risk*

Supplementary material S3 shows the prediction and classification instability plots had the existing dataset only had 221 participants, the minimum required to estimate the overall risk precisely (criterion (i)). The uncertainty intervals are now extremely wide and the classification instability index is quite large for most individuals.

5. Results II: Examination of stability in subgroups to inform model fairness checks

Robustness of model predictions in subgroups may form part of fairness checks for using a model in practice, and is a recommendation within the TRIPOD+AI reporting guideline.[14] Thus, as part of the sample size calculations, it may be important to examine anticipated uncertainty intervals and classification instability in subgroups defined by relevant patient characteristics. For this reason, in Step (1) we mentioned that the core predictors may include variables that represent protected characteristics. The model might include such predictors in the 'core model' or leave them out; regardless, precision and classification instability can be checked as long as the relevant variables are available after Step (1). To illustrate this, we return to the acute kidney injury example and examine subgroups defined by ethnicity.

In the original dataset, ethnicity was recorded as a categorical variable, with the following distribution: white (72%), black (8%), Asian (2%), Hispanic (4%), other (3%) and unknown (11%). In the full dataset of 20413 participants, this corresponds to 14774 white, 1584 black, 508 Asian, 731 Hispanic, 579 other and 2237 unknown. Hence, although the overall sample size and number of whites is very large, some other ethnic groups are relatively low. This is even more evident in a random sample of 511 participants (the minimum recommended by *pmsampsize*), with about 370 white, 40 black, 13 Asian, 18 Hispanic, 14 other and 56 unknown. This may raise concerns that any model would be unreliable and imprecise in, for example, black, Asian and Hispanic groups.

***Ethnicity not included in the 'core model'***

Recall, ethnicity was not an included predictor in the 'core model'; this will sometimes be deemed appropriate in 'fear of making clinical decisions race-sensitive'.[40] Thus, the small sample sizes for some ethnic groups is not relevant in terms of estimating the prognostic effect of ethnicity itself. Rather, the imprecision of risk estimates for each ethnic group will depend on their relationship with the joint distribution of predictors actually included in the 'core model'. Our sample size approach helps examine this, by examining the expected uncertainty interval widths and classification instability (based on a 10% risk threshold) for each ethnic group, as their relationship with the core predictors is embedded in the existing dataset containing the core predictors.

Reassuringly, regardless of the specified total sample size, the precision of estimated risks and classification stability appears similar across ethnic groups. For example, with a total sample size of 511 participants, the mean (min, max) uncertainty interval widths are 0.19 (0.0065, 0.77) for whites and 0.17 (0.0048, 0.57) for Asians; the average (min, max) MAPE is 0.039 (0.0012, 0.18) for whites and 0.035 (0.00082, 0.12) for Asians; and the average (min, max) probability of misclassification of 0.083 (0, 0.50) for whites and 0.070 (0, 0.49) for Asians.

***Ethnicity included in the 'core model'***

Now, let us consider if ethnicity *had* been included as one of the core predictors in the 'core model' and that non-white ethnic groups have a higher risk than whites. Whites are the reference group, and the other five ethnic groups have the same relative weight as other core predictors included in the model. With the additional five parameters, *pmsampsize* now suggests a minimum sample size of 795 participants. At this sample size, the expected uncertainty interval widths, MAPE and misclassification probabilities are now quite discrepant across ethnic groups (Table 1, Figure 6); for example, about twice as high for Asians as for whites. This issue is masked when looking at all participants combined because, as the majority of the population are white, the summary results for all participants are similar to whites alone (Table 1).

*Table 1: Summary statistics for the expected precision and probability of misclassification for the AKI model when developed using 795 participants*

| Ethnicity | 95% uncertainty interval width: mean (min, med, max) | Mean Absolute Prediction Error (MAPE): mean (min, med, max) | Probability of misclassification: mean (min, med, max) |
|---|---|---|---|
| Asian | 0.36 (0.023, 0.37. 0.62) | 0.074 (0.0040, 0.075, 0.14) | 0.14 (0, 0.089, 0.50) |
| Black | 0.27 (0.016, 0.27, 0.67) | 0.055 (0.0025, 0.056, 0.15) | 0.089 (0, 0.0097, 0.50) |
| Hispanic | 0.31 (0.032, 0.31, 0.65) | 0.064 (0.0063, 0.062, 0.15) | 0.14 (0, 0.078, 0.50) |
| Other | 0.35 (0.035, 0.35, 0.65) | 0.072 (0.0066, 0.071, 0.15) | 0.14 (0, 0.089, 0.50) |
| Unknown | 0.23 (0.018, 0.23, 0.60) | 0.047 (0.0033, 0.045, 0.13) | 0.086 (0, 0.0094, 0.50) |
| White | 0.15 (0.012, 0.13, 0.68) | 0.031 (0.0022, 0.027, 0.15) | 0.082 (0, 0.0058, 0,50) |
| *Overall* | 0.19 (0.012, 0.16, 0.68) | 0.038 (0.0022, 0.032, 0.15) | 0.088 (0, 0.0097, 0.50) |

Unlike when ethnicity was *not* included in the model, discrepancies occur here because the prognostic effect of each ethnic group is now estimated during model development, and the uncertainty in these estimates is propagated when making predictions for each group. As the sample size is small (795 participants) and a large proportion of the population are white, the non-white ethnic groups have quite sparse data and their estimated risks are far more imprecise than for whites (Figure 6).

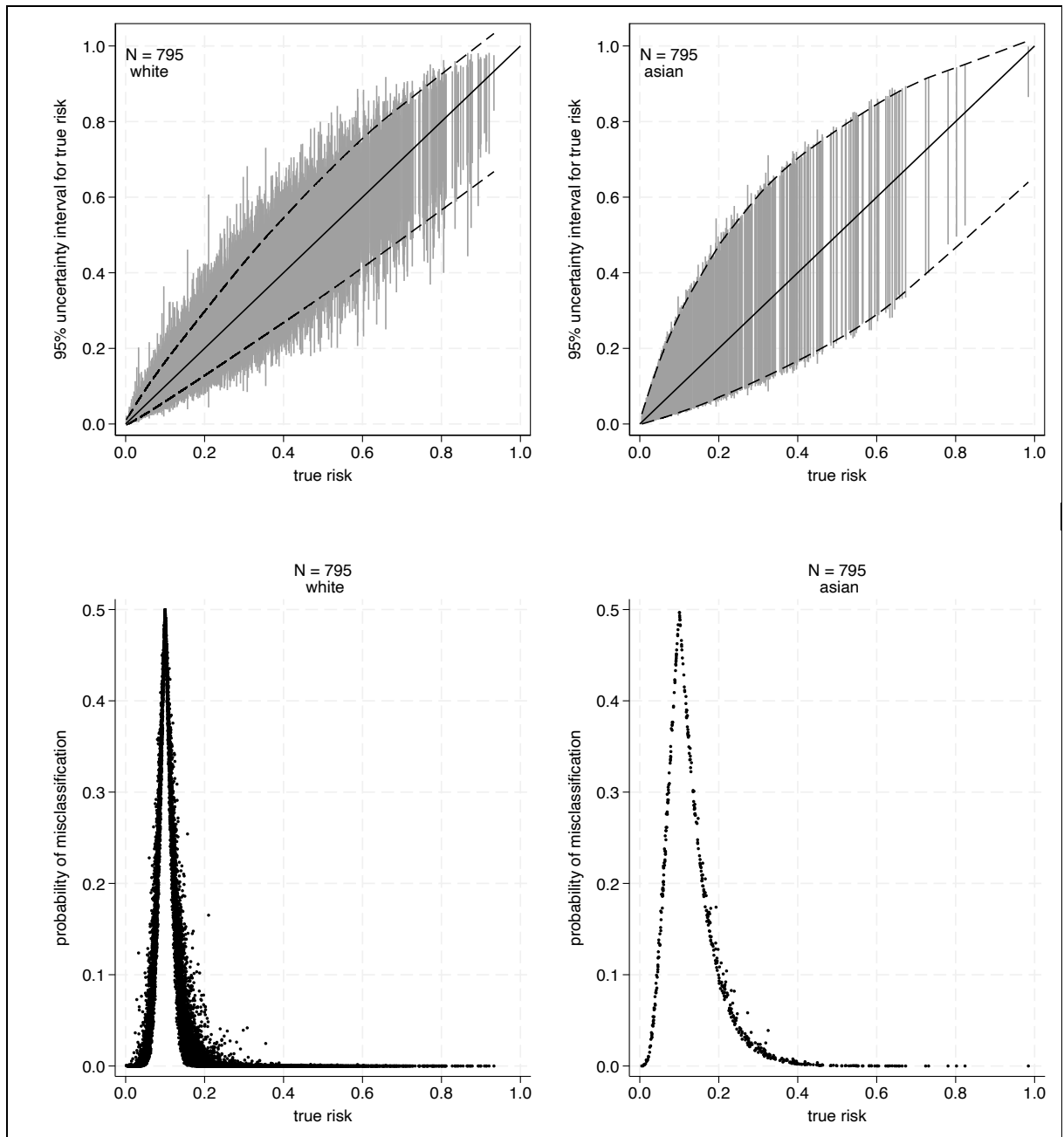

*Figure 6 Prediction instability (first row) and classification instability (bottom row) plots for whites (first column) and Asians (second column) in the full dataset for the acute kidney injury model when using 795 participants to develop the model, and assuming a 10% risk threshold, with ethnicity included in the model*

This demonstrates a tension about developing a model including a predictor with categories that are sparsely represented in the development dataset: though the predictor may help improve discrimination performance, it may lead to larger uncertainty of risk estimates for participants in the sparse category. Our sample size approach can help expose this in advance of model development and instigate discussion with stakeholders (e.g. representatives from each ethnic group) about whether any projected discrepancies across subgroups in uncertainty interval widths and misclassification probabilities are unfair. If so, larger sample sizes are needed to address this. Indeed, in the original dataset of 20413 participants, unfairness concerns are diminished due to the much larger sample size.

In situations where categories of a predictor are more evenly distributed in the target population, discrepancies in uncertainty of risk estimates will be smaller. For example, supplementary material S4 shows similar uncertainty for males and females, as these were similarly represented (females 40%, males 60%).

6. Results III: Evaluation and comparison to other machine learning methods

Our proposal is theoretically driven by Fisher's information matrix based on unpenalised logistic regression models; it is thus based on exactly the standard errors of individual predictions that would be obtained from a fitted logistic regression model with the same unit information matrix as assumed for the 'core model'. However, the question remains whether the approach is still relevant for other model development methods are used. Further research is needed, but we make the following initial recommendations.

- Criteria (ii) and (iii) target a small amount of overfitting, and in this situation unpenalised logistic regression will be quite similar to when using penalised regression approaches (e.g., uniform shrinkage, lasso, elastic net and ridge regression).
- We deliberately recommend to only consider core predictors in our new uncertainty-based approach. We recognise that (many) other predictors might be considered during model building, but this will typically add further instability that then would require larger sample sizes. Hence, our approach based on a small set of (strong) predictors is a minimum sample size calculation, in keeping with our outlined criteria (i) to (iii).

- Sample size for other machine learning approaches, such as tree-based methods, often need substantially higher sample sizes to achieve the same level of stability compared to (penalised) regression approaches.[41] This is typically the case when using the default settings in software packages, as they tend to allow for large complexity (e.g., large numbers of trees).

We illustrate these points in supplementary material S5 for the foot ulcer and kidney examples.

7. Discussion

The development of a reliable clinical prediction model is dependent on the sample and size of data used to develop it; were a different sample of the same size used from the same overarching population, the developed model and subsequent predictions could be very different.[20, 22] This issue is rarely considered in clinical prediction model research, as most models just provide a single predicted estimate for each individual, and any uncertainty of that estimate is largely ignored. In this article, we have proposed that researchers should formally consider this issue in advance of model development during the design of the study, by examining the sample size required to produce individual risk estimates that are acceptably precise for the clinical context at hand. This can be done prior to collecting new data, or when deciding if an existing dataset is suitably large for model development, perhaps as part of a grant application. Two examples showcased the approach in action, and our software *pmstabilityss* implements the approach.

We also emphasised the importance of considering the actual clinical setting of interest, with input from key stakeholders including health professionals and patients about acceptable uncertainty intervals and, if relevant, classification instability. Currently, such discussions between the model developers and stakeholders are either not done at all or tend to be conducted after model development. We hope our approach will force the communication to begin at the onset of the development study, in the context of the clinical decision to be made. Where risk thresholds are relevant, this can be framed within a decision-theoretic perspective. Sadatsafavi et al examine this at the population-level, in regard to value of information and net benefit.[42, 43] In our approach, we consider classification instability (probability of misclassification) at the individual level. Further research is needed about this individual-level decision-analysis premise and value of information investigations. For instance, in our examples we focused on one or two thresholds typical of people in the population, but a fully individual-level approach would allow unique thresholds for each individual. This may be difficult to implement for a sample size calculation. Also, risk thresholds are not always relevant; the aim may simply be to ensure that predictions are precise for most individuals, and so then other metrics (e.g., ensuring the majority of uncertainty interval widths are less than, say, 0.1) are more relevant.

Our proposal utilises maximum likelihood estimation theory for standard (unpenalised) logistic regression models, and formally examines epistemic uncertainty (*reducible* model-based uncertainty) that arises from fitting a logistic regression with a core set of predictors.[44] We do not consider aleatoric uncertainty (*irreducible* uncertainty) that refers to residual uncertainty that cannot be explained by the 'core model'. A key contribution is to propose a closed-form solution for the variance of individual-level predictions that is decomposed into Fisher's unit information matrix and the total sample size. This makes the computation relatively fast as, once the unit information matrix is derived, variances of individual-level risk estimates (and subsequent uncertainty intervals) can be quickly calculated for a range of sample sizes to examine the impact on uncertainty interval widths and classification instability. If additional predictors to the core set are of interest, then this is likely to inflate the uncertainty further, and so the sample size required for the core set is a minimum. Using the framework of Heinze et al.,[45] we view our approach as Phase 3 methodology work for those building (penalised) logistic regression models in situations where overfitting is anticipated to be low. For other machine learning approaches and settings with small data and large numbers of predictors, further research is needed building on Section 6.[46, 47]

Our work is also aiming to improve fairness of developed prediction models, as it allows researchers to plan and target sufficiently precise predictions for key subgroups and those with protected characteristics. Section 5 highlighted this issue when ethnicity was included in the 'core model'. Such characteristics can be checked even when not included in the 'core model', especially as it may be pertinent to exclude certain variables (e.g., ethnicity) from models if their predictive effects propagate bias. We recognise precision and instability checks do not cover all aspects of fairness, and that precise risk estimates do not necessarily lead to health equity.

The main obstacles to our approach are specifying the 'core model' and the joint predictor distributions. Choices inevitably involve some subjectivity, and so sensitivity analyses may be sensible, for example by examining how uncertainty intervals change for other plausible 'core model' specifications (e.g. different predictor weights; allowing for non-linear associations; assuming independent or correlated predictors). Nevertheless, we provided pragmatic suggestions to facilitate the process, for example by focusing on a small number of core predictors (and protected characteristics) of interest, and by basing the 'core model' on previously published models (see our Example 1) or by assuming equal weighting of (standardised) predictors whilst adhering to a particular overall risk and C-statistic (see our Example 2). Generally, our approach is more easily implemented when an existing dataset is available (which is often the situation in practice), as then the joint distribution of core predictors can be observed directly. In situations in advance of data collection, the joint distribution of predictors may be difficult to gauge and assuming predictors are independent may be a pragmatic approach; the impact of that needs further research but it forms a starting point. Then, as new data are being collected and joint distributions become observable (e.g. in a pilot study), the sample size calculation can be updated.

In summary, we have proposed an uncertainty-based sample size calculation for developing a clinical prediction model for a binary outcome. The approach enables researchers to examine how the sample size (for new data collection or an existing dataset) impacts individual-level uncertainty intervals and classification instability, to guide decisions on suitable datasets and sample size targets for model development. A companion article (part 2) will extend to time-to-event outcomes and datasets with censoring.

57. Kaplan J. Decision Theory and the Factfinding Process. Stanford Law Review. 1968;20(6):1065-92.

# SUPPLEMENTARY MATERIAL

## S1: Further details of existing sample size calculation

Criterion (i): sample size to target a precise estimate of the overall outcome risk

Precision in the overall risk is the first stability level defined by Riley and Collins.[20] If the sample size is not even large enough to estimate the overall risk precisely at the population level, then it is futile to consider estimating individual-level risks. For a binary outcome, based on the anticipated overall risk ($\hat{\phi}$) and a target absolute margin of error ($\delta$), corresponding to a target 95% confidence interval width of $2\delta$, the minimum required sample size ($n$) to estimate the overall risk precisely is approximately:[15]

$$n = \left[ \left( \frac{1.96}{\delta} \right)^2 \hat{\phi}(1 - \hat{\phi}) \right] \quad \text{Eq. (10)}$$

We generally recommend aiming for $\delta \leq 0.05$, and thus confidence interval width $\leq 0.1$ but a smaller margin of error may be sensible for low (or high) event risks.

Criterion (ii): sample size to target small overfitting of predictor effects

Shrinkage (also known as penalisation or regularisation) methods aim to address the problem of overfitting by reducing the variability in a developed model's predictions such that otherwise extreme predictions (i.e., predicted risks closest to 0 or 1) are pulled towards the overall average.[48-53] However, there is no guarantee that shrinkage or penalisation methods will fully overcome the problem of overfitting;[46, 49] and the larger the shrinkage required, the greater concern that model predictions will be miscalibrated.

To address this, our second criterion targets a sample size ($n$) and number of candidate predictor parameters ($P$) that minimise the problem of overfitted predictor effects.[15, 16, 54] The calculation is based on theory for regression models, and requires the researcher to pre-specify the number of candidate predictor parameters ($P$), a uniform shrinkage factor ($S$, which we recommend to be $\leq 0.9$ so that overfitting is $\leq 10\%$), and the anticipated model performance as defined by the Cox-Snell R-squared statistic ($R^2_{CS}$).[15, 50, 55] The minimum sample size ($n$) is then:

$$n = \left\lceil \frac{P}{(S-1)\ln\left(1 - \frac{R_{CS}^2}{S}\right)} \right\rceil \qquad \text{Eq. (11)}$$

The number of parameters ($P$) should, at least, correspond to a core set of predictors known to be important in the field (see Section 3 for more discussion on this). Anticipated values of $R_{CS}^2$ can be taken from previous studies (existing models in the same field) or, in the absence of any other information, assuming the value of $R_{CS}^2$ corresponds to an $R_{Nagelkerke}^2$ of 0.15 (i.e. $R_{CS}^2 = 0.15 \times max(R_{CS}^2)$, where $\max(R_{CS}^2) = 1 - \left(\phi^\phi (1-\phi)^{1-\phi}\right)^2$)), such that 15% of the total variance is explained.[15] The value of $R_{CS}^2$ can also be derived from a specified $C$ statistic.[25]

Criterion (iii): sample size to target a small optimism in apparent model fit

The final criterion targets a small difference in the developed model's apparent and optimism-adjusted values of $R_{Nagelkerke}^2$ (= $R_{CS}^2/\max(R_{CS}^2)$), as this is a fundamental overall measure of model fit.[2, 56] The apparent $R_{Nagelkerke}^2$ value is the model's observed performance in the same data used to develop the model, whilst the optimism-adjusted $R_{Nagelkerke}^2$ value is a more realistic (approximately unbiased) estimate of the model's fit in the target population. The approach calculates the shrinkage factor that corresponds to an expected optimism of $\delta$ in $R_{Nagelkerke}^2$:[15]

$$S = \frac{R_{CS}^2}{R_{CS}^2 + \delta \max(R_{CS}^2)} \qquad \text{Eq. (12)}$$

We suggest $\delta$ is a small value, such as $\leq 0.05$. The obtained value of $S$ can then be placed into the previous equation from step 2, to calculate the minimum required sample size ($n$)

## S2: Explanation of how risk thresholds are chosen within a decision-theory perspective

It is helpful to formalise the choice of risk thresholds within a decision analysis framework. Assume that decision is whether to biopsy (two possible actions: yes or no) and there are two possible states (prostate cancer present: yes or no), which creates four possible scenarios (pathways) shown in Figure S1. Assigned to each scenario is a "utility" (U1 to U4), which is a numerical measure that defines a value placed on a given pathway. These are specific to an individual, and essentially measure their preference for each scenario if it were known that the assumed state in that scenario was correct. The values of U1 to U4 are best considered relative to one another. For example, consider an individual expresses their utility of each pathway as: $U1_i = 100$, $U2_i = 5$, $U3_i = 0$ and $U4_i = 10$; this means they are expressing the action of biopsy if they do have prostate cancer ($U1_i = 100$), to be 10 times more important than the action of no biopsy if they do not have prostate cancer ($U4_i = 10$).

*Figure S1 Summarising the four possible pathways that stem from the decision of whether or not to request a biopsy in an individual that may or may not have prostate cancer*

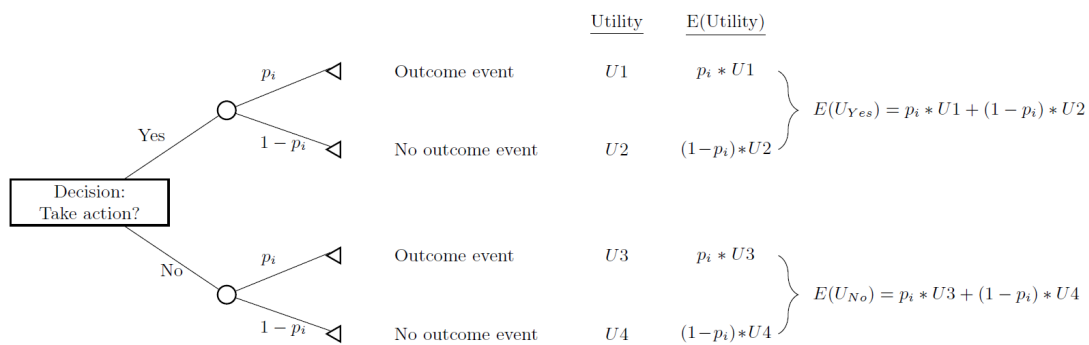

The individual's chosen utilities also define their risk threshold at which they would be willing to choose a biopsy. It corresponds to where the expected utility from biopsy ($E(U_{Biopsy})$) exceeds the expected utility from no biopsy (e.g., $E(U_{NoBiopsy})$):

$$E(U_{Biopsy}) > E(U_{No\ Biopsy})$$

This can be expressed in terms of the individual's probability of prostate cancer ($p_i$) weighted by the utility values:

$$(p_i * U1_i) + ((1 - p_i) * U2_i) > (p_i * U3_i) + ((1 - p_i) * U4_i)$$

Rearranging identifies the risk threshold at which the individual prefers a biopsy.[57]

$$p_{\text{THRESHOLD}i} > \left(1 + \frac{U1_i - U3_i}{U4_i - U2_i}\right)^{-1}$$

For example, let us return to the individual who expresses their utility of each outcome state as $U1_i = 100$, $U2_i = 5$, $U3_i = 0$ and $U4_i = 10$. Then:

$$p_{\text{THRESHOLD}i} > \left(1 + \frac{U1_i - U3_i}{U4_i - U2_i}\right)^{-1} = \left(1 + \frac{100 - 0}{10 - 5}\right)^{-1} = 0.048$$

If a model estimates this individual's risk of prostate cancer to be $\hat{p}_i = 0.051$, then this suggests their preference is a biopsy, as their point estimate of risk exceeds their personal risk threshold, and so their expected utility of a biopsy is larger than their expected utility of no biopsy:

$$E(U_{Biopsy}) = (\hat{p}_i * U1_i) + ((1 - \hat{p}_i) * U2_i) = (0.051 * 100) + ((1 - 0.051) * 5) = 9.85$$

$$E(U_{No\ Biopsy}) = (\hat{p}_i * U3_i) + ((1 - \hat{p}_i) * U4_i) = (0.051 * 0) + ((1 - 0.051) * 10) = 9.49$$

Even though this suggests the correct decision is to biopsy, there may still be uncertainty about this decision depending on the precision of $\hat{p}_i$. Indeed, understanding the prediction uncertainty may give doctors, health professionals and regulators assurance to use or endorse the model in the first place, or identify when further research is needed. Vickers et al. support this argument,[33] noting that "decision analysis tells us which decision to make for now, but we may also want to know how much confidence we should have in that decision. If we are insufficiently confident that we are right, further research is warranted." In this context, the aim of our sample size approach is to help understand and examine which sample sizes are likely to give sufficient information to guide decisions.

## S3: Instability for the acute kidney model with 212 participants (the minimum required to estimate the overall risk precisely)

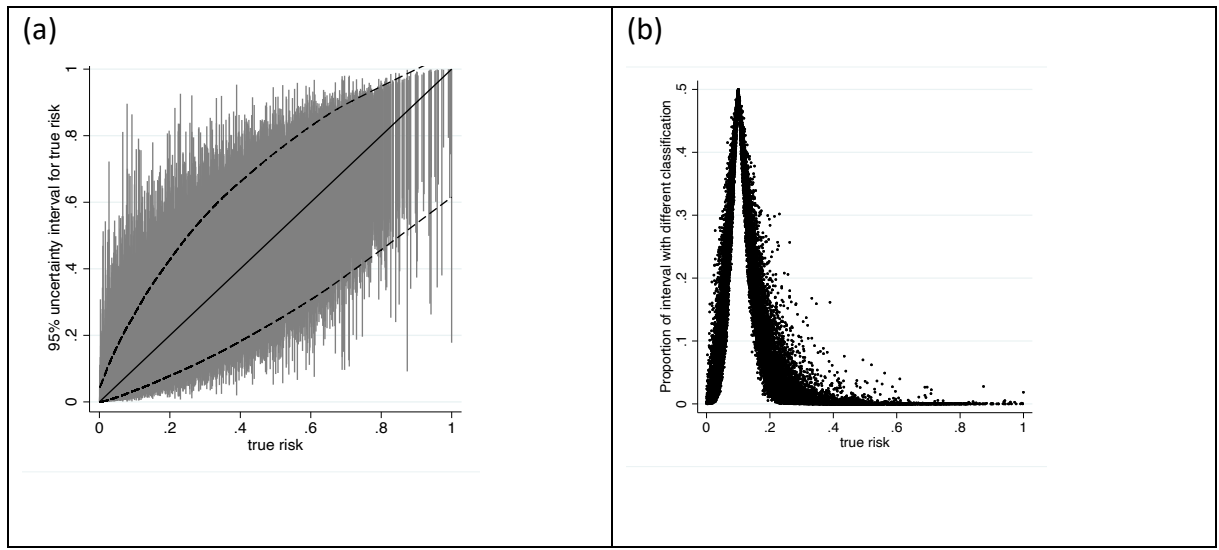

*Figure S2: Expected uncertainty interval widths and classification instability when developing an acute kidney injury prediction model with a sample size of 212 participants, the minimum required to estimate the overall risk precisely*

## S4: Instability by sex for the acute kidney injury model

Recall that sex was a predictor in the 'core model'. Previously we examined prediction and classification instability plots across all individuals, but now let us consider males and females separately, and assume a risk threshold of 10% is of interest for everyone. Regardless of whether we consider the full sample size of 20413 participants (12186 males, 8227 females) or the smaller sample size of 511 participants (305 males, 206 females), the width of uncertainty intervals and magnitude of classification instability appear quite similar for males and females (figures shown below). Hence, there does not seem to be any strong concerns of a discrepancy in the model's precision or robustness based on sex, in this example.

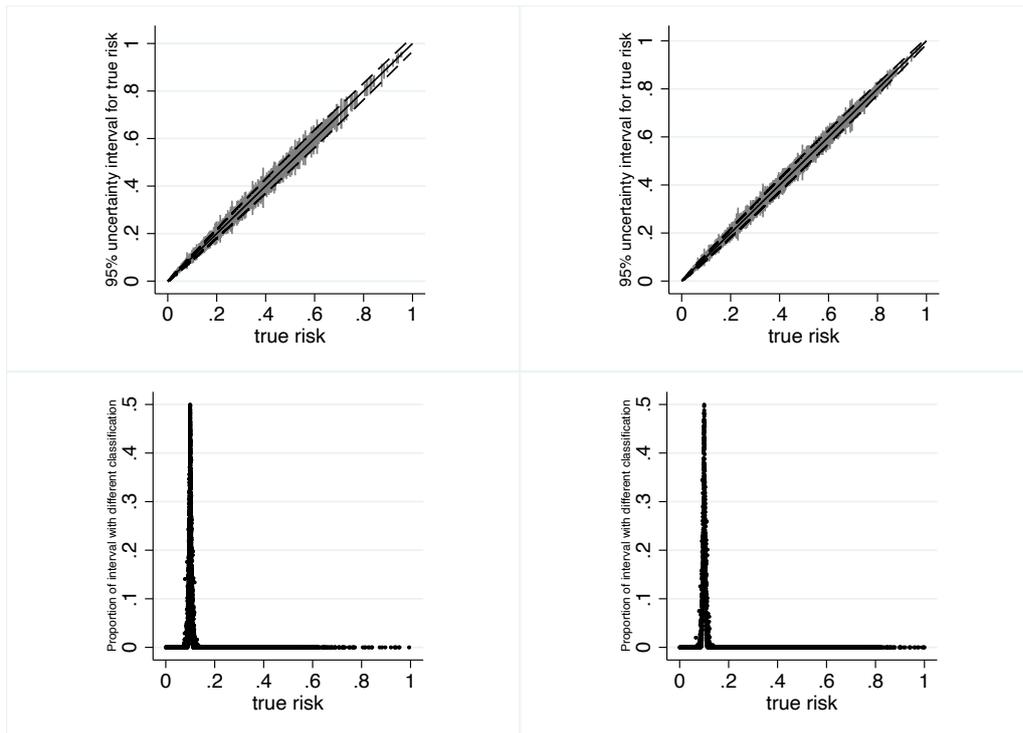

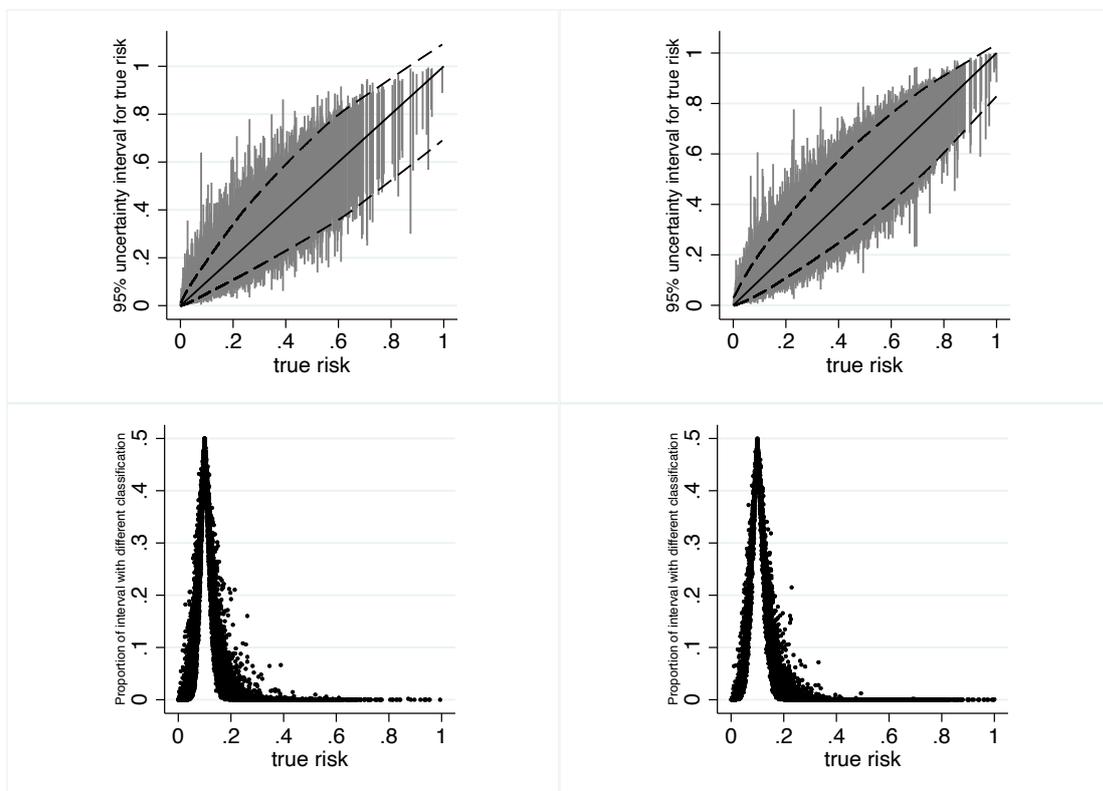

*Figure S3: Prediction and classification instability plots by sex for the acute kidney injury model with 20431 and 511 participants*

**S5: Comparison of uncertainty intervals for various modelling approaches**

(i)  Uncertainty intervals for the Chappell example for various modelling approaches

Let us return to the Chappell example and consider empirically-based uncertainty intervals from a model developed using a random forest, rather than logistic regression, for the identified sample size of 1224 participants. To devise the empirical-based uncertainty interval, we developed 1000 models each trained on three predictors and a random sample of 1224 participants (sampled from the full synthetic data generated in steps 1 to 3 in section 4), with outcome values also randomly generated for each individual based on their assumed 'true' risk as specified by the core model in Step 3. Each forest was allowed 100 trees with tree depth of 3, and we applied each of the 1000 models to each participant in the full synthetic dataset to obtain 1000 estimated risks for every individual. The 2.5% and 97.5% percentile values were then used to derive a 95% uncertainty interval for each individual's risk, as shown in Figure(a). These are considerably wider than the intervals based on the logistic regression (Figure (b)). Varying the number of trees did not affect the findings in this example. For comparison, the empirical-based uncertainty intervals are shown for an unpenalised logistic regression model in Figure (b), which are identical to those from our sample size approach (Figure (d)), as should be anticipated given the theory is based on this approach. The intervals are also very similar to those from a logistic regression with a lasso penalty (Figure(c)), again as might be expected given that criteria (ii) and (iii) aim to minimise overfitting such that penalised and unpenalised approaches are similar.

Next, compare empirically-based uncertainty intervals for the acute kidney injury example and various model development approaches (Figure S4). Again, the interval widths from an unpenalised logistic regression are practically identical to those from our sample size calculation, as expected, and those from the lasso are also very similar. The intervals from the random forest are much larger when using the default software settings for number of trees and depth. There also appears to be concerns of miscalibration.

| (a) random forest | (b) unpenalised logistic regression |

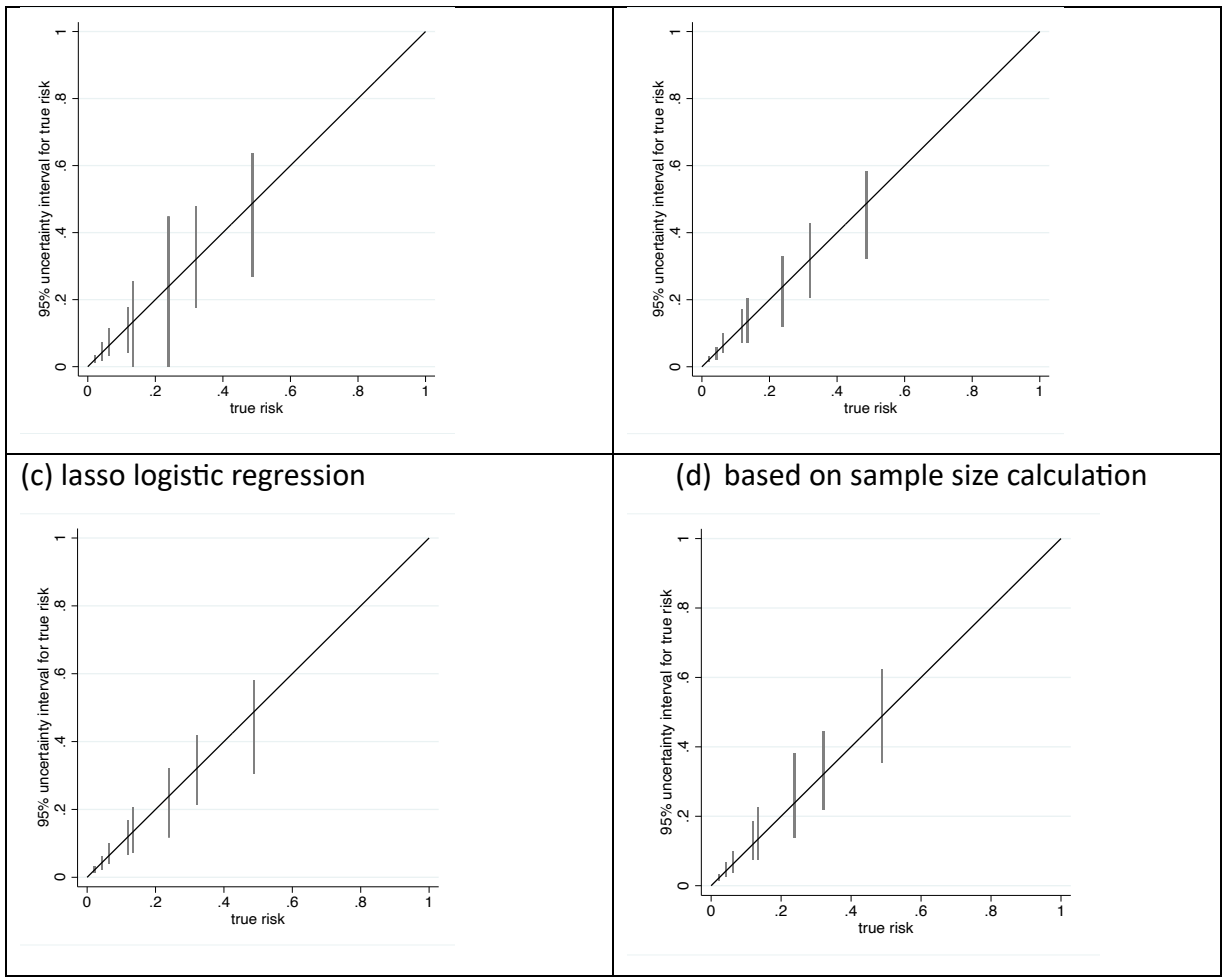

*Figure S4: Empirical-based uncertainty intervals for individual risks from 1000 models of foot ulcer risk developed using 1224 participants and either (a) a random forest (100 trees and depth of 3), (b) an unpenalised logistic regression, and (c) a logistic regression with lasso penalty. Uncertainty intervals are obtained from 2.5% and 97.5% percentiles of each individual's 1000 predictions from the 1000 models. For comparison, panel (d) shows the uncertainty intervals derived from our sample size calculation based on the same unit information as the fitted models, and thus give identical results to panel (b).*

**(ii)    Uncertainty intervals for the kidney example for various modelling approaches**

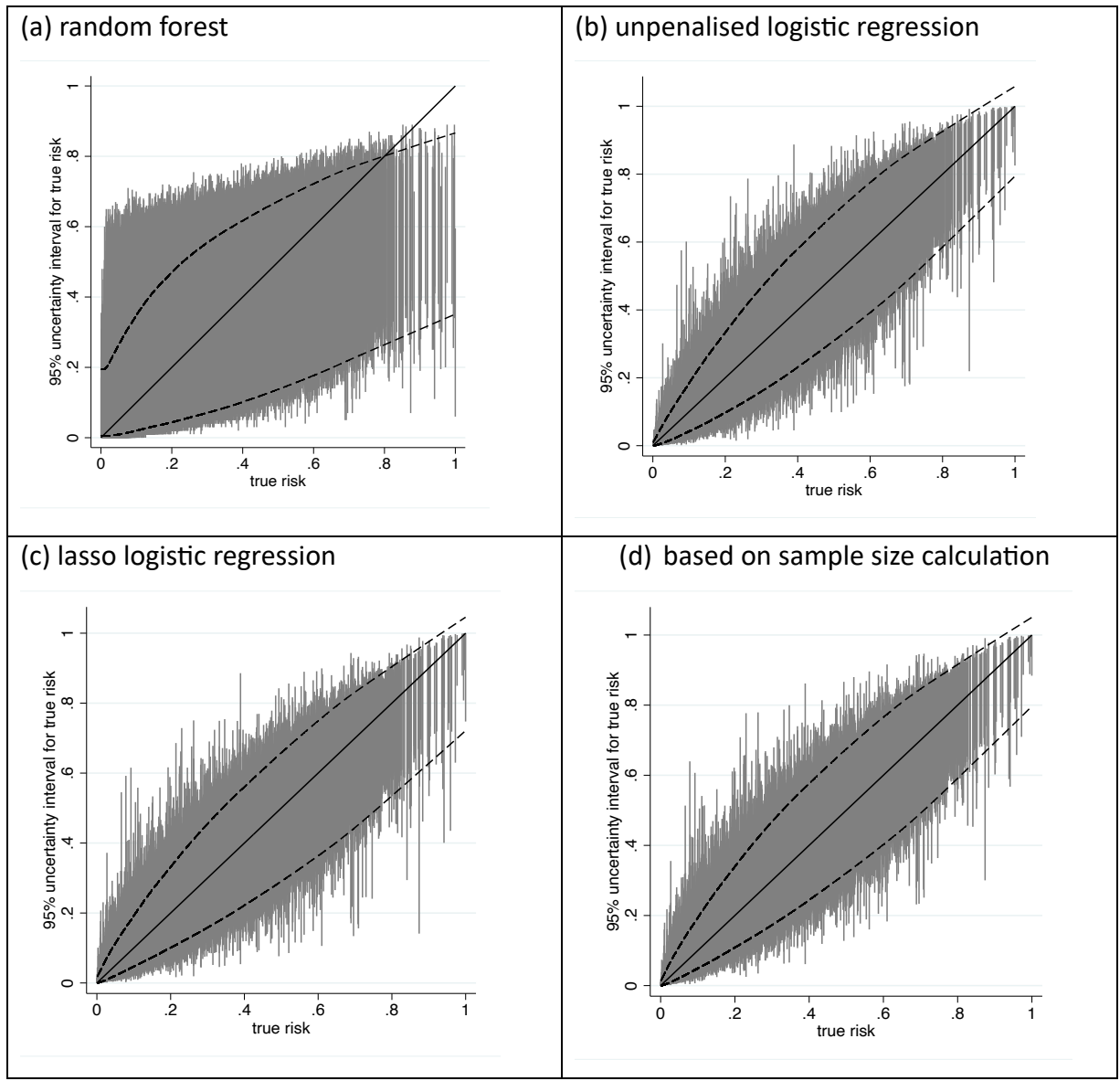

*Figure S5: Empirical-based uncertainty intervals for individual risks from 1000 models of acute kidney injury risk developed using 511 participants and either (a) a random forest (default software settings), (b) an unpenalised logistic regression, and (c) a logistic regression with lasso penalty Uncertainty intervals are obtained from 2.5% and 97.5% percentiles of each individual's 1000 predictions from the 1000 models. For comparison, (d) shows the uncertainty intervals derived from our sample size calculation.*